\documentclass[preprint2]{aastex}

\newcommand{\kms}{\ensuremath{\rm km\,s^{-1}}}
\newcommand{\ms}{\ensuremath{\rm m\,s^{-1}}}
\newcommand{\gcmc}{\ensuremath{\rm g\,cm^{-3}}}

\newcommand{\teff}{\ensuremath{T_{\rm eff}}}
\newcommand{\logg}{\ensuremath{\log{g}}}
\newcommand{\vsini}{\ensuremath{v \sin{i}}}
\newcommand{\feh}{[Fe/H]}

\newcommand{\rsun}{\ensuremath{R_\sun}}
\newcommand{\msun}{\ensuremath{M_\sun}}
\newcommand{\lsun}{\ensuremath{L_\sun}}

\newcommand{\rstar}{\ensuremath{R_\star}}
\newcommand{\mstar}{\ensuremath{M_\star}}
\newcommand{\loggstar}{\ensuremath{\logg_\star}}
\newcommand{\lstar}{\ensuremath{L_\star}}

\newcommand{\rpl}{\ensuremath{R_{\rm P}}}
\newcommand{\mpl}{\ensuremath{M_{\rm P}}}

\newcommand{\rhopl}{\ensuremath{\rho_{\rm P}}}
\newcommand{\loggpl}{\ensuremath{\logg_{\rm P}}}
\newcommand{\teq}{\ensuremath{T_{\rm eq}}}

\newcommand{\rjup}{\ensuremath{R_{\rm J}}}
\newcommand{\mjup}{\ensuremath{M_{\rm J}}}

\newcommand{\koicur}{Kepler-4}
\newcommand{\koicurb}{Kepler-4b}

\newcommand{\koicurCCra}{\ensuremath{19^{\mathrm{h}}02^{\mathrm{m}}27^{\mathrm{s}}.68}}
\newcommand{\koicurCCdec}{\ensuremath{+50^{\circ}08'08''.7}}
\newcommand{\koicurCCkic}{KIC~11853905}
\newcommand{\koicurCCtwomass}{2MASS~19022767+5008087}
\newcommand{\koicurCCkicr}{12.211}			

\newcommand{\koicurLCar}{\ensuremath{6.47^{+0.26}_{-0.28}}}			
\newcommand{\koicurLCrprstar}{\ensuremath{0.02470^{+0.00031}_{-0.00030}}}	
\newcommand{\koicurLCimp}{\ensuremath{0.022^{+0.234}_{-0.022}}}			
\newcommand{\koicurLCi}{\ensuremath{89.76^{+0.24}_{-2.05}}}			

\newcommand{\koicurLCP}{\ensuremath{3.21346\pm0.00022}}	
\newcommand{\koicurLCPshort}{3.213}				
\newcommand{\koicurLCPprec}{\ensuremath{3.21346}}
\newcommand{\koicurLCT}{\ensuremath{2454956.6127\pm0.0015}}	
\newcommand{\koicurLCdurshort}{\ensuremath{3.95}}

\newcommand{\koicurSMEteff}{\ensuremath{5857\pm120}}	
\newcommand{\koicurSMEfeh}{\ensuremath{+0.17\pm0.06}}	
\newcommand{\koicurSMElogg}{\ensuremath{4.25\pm0.10}}	
\newcommand{\koicurSMEvsin}{\ensuremath{2.2\pm1.0}}	

\newcommand{\koicurYYmlong}{\ensuremath{1.223^{+0.053}_{-0.091}}}	%
\newcommand{\koicurYYrlong}{\ensuremath{1.487^{+0.071}_{-0.084}}}	%
\newcommand{\koicurYYlogg}{\ensuremath{4.17\pm0.04}}			%
\newcommand{\koicurYYlum}{\ensuremath{2.26^{+0.66}_{-0.48}}}		%
\newcommand{\koicurYYmv}{\ensuremath{4.00\pm0.28}}			%
\newcommand{\koicurYYage}{\ensuremath{4.5\pm1.5}}			%
\newcommand{\koicurRVK}{\ensuremath{9.3^{+1.1}_{-1.9}}}			
\newcommand{\koicurRVgamma}{\ensuremath{-1.27\pm1.1}}			
\newcommand{\koicurRVmean}{\ensuremath{-61.0\pm0.10}}			

\newcommand{\koicurPPlogg}{\ensuremath{3.16^{+0.06}_{-0.10}}}			%
\newcommand{\koicurPParel}{\ensuremath{0.0456\pm0.0009}}		
\newcommand{\koicurPPrho}{\ensuremath{1.91^{+0.36}_{-0.47}}}		%
\newcommand{\koicurPPrhoshort}{\ensuremath{1.9}}			%

\newcommand{\koicurPPm}{\ensuremath{0.077\pm0.012}}	 		%
\newcommand{\koicurPPmlong}{\ensuremath{0.077\pm0.012}}	%
\newcommand{\koicurPPmEarth}{\ensuremath{24.5\pm3.8\ M_{\earth}}}
\newcommand{\koicurPPr}{\ensuremath{0.357\pm0.019}}			%
\newcommand{\koicurPPrlong}{\ensuremath{0.357\pm0.019}}			%
\newcommand{\koicurPPrEarth}{\ensuremath{3.99\pm0.21\ R_{\earth}}}
\newcommand{\koicurPPteq}{\ensuremath{1650\pm200}}			%
\newcommand{\koicurXdist}{\ensuremath{550\pm80}}			%

\shortauthors{Borucki et al.}
\shorttitle{\koicurb}

\begin{document}


\title{Kepler-4b: Hot Neptune-Like Planet of a G0 Star Near Main-Sequence
Turnoff\altaffilmark{\dagger}}

\altaffiltext{$\dagger$}{Some of the data presented herein were obtained at the W.M. Keck Observatory, which is operated as a scientific partnership among the California Institute of Technology, the University of California and the National Aeronautics and Space Administration. The Observatory was made possible by the generous financial support of the W.M. Keck Foundation.}


\author{
William~J. Borucki,\altaffilmark{1}
David~G. Koch,\altaffilmark{1}
Timothy~M. Brown,\altaffilmark{2}
Gibor Basri,\altaffilmark{3}
Natalie~M. Batalha,\altaffilmark{4}
Douglas~A. Caldwell,\altaffilmark{1}
William~D. Cochran,\altaffilmark{5}
Edward~W. Dunham,\altaffilmark{7}
Thomas~N. Gautier III,\altaffilmark{9}
John~C. Geary,\altaffilmark{8}
Ronald~L. Gilliland,\altaffilmark{12}
Steve~B. Howell,\altaffilmark{11}
Jon~M. Jenkins,\altaffilmark{6,1}
David~W. Latham,\altaffilmark{8}
Jack~J. Lissauer,\altaffilmark{1}
Geoffrey~W. Marcy,\altaffilmark{3}
David Monet,\altaffilmark{10}
Jason~F. Rowe, \altaffilmark{1} and
Dimitar Sasselov\altaffilmark{8}
}
\altaffiltext{1}{NASA/Ames Research Center, Moffett Field, CA 94035}
\altaffiltext{2}{Las Cumbres Observatory Global Telescope, Goleta, CA, 93117}
\altaffiltext{3}{University of California-Berkeley, Berkeley, CA 94720}
\altaffiltext{4}{San Jose State University, San Jose, CA 95192}
\altaffiltext{5}{University of Texas at Austin, Austin, TX 78712}
\altaffiltext{6}{SETI Institute, Mountain View, CA 94043}
\altaffiltext{7}{Lowell Observatory, Flagstaff, AZ 86001}
\altaffiltext{8}{Harvard-Smithsonian Center for Astrophysics, Cambridge, MA 02138}
\altaffiltext{9}{Jet Propulsion Laboratory/California Institute of Technology, Pasadena, CA 91109}
\altaffiltext{10}{United States Naval Observatory, Flagstaff, AZ 86002}
\altaffiltext{11}{National Optical Astronomy Observatories, Tucson, AZ 85726}
\altaffiltext{12}{Space Telescope Science Institute, Baltimore, MD 21218}









\begin{abstract}
Early time-series photometry from NASA's {\it Kepler} spacecraft has revealed a
planet transiting the star we term Kepler-4, at $RA = \koicurCCra$,
$\delta = \koicurCCdec$.
The planet has an orbital period of $\koicurLCPshort$ days and shows transits with
a relative depth of $0.87 \times 10^{-3}$ and a duration of about
$\koicurLCdurshort$ hours.
Radial velocity measurements from the Keck HIRES spectrograph show a
reflex Doppler signal of $\koicurRVK$ $\ms$, consistent with a low-eccentricity
orbit with the phase expected from the transits.
Various tests show no evidence for any companion star near enough to
affect the light curve or the radial velocities for this system.
From a transit-based estimate of the host star's mean density, combined
with analysis of high-resolution spectra, we infer that the host star
is near turnoff from the main sequence, with estimated mass and radius of
\koicurYYmlong ~\msun and \koicurYYrlong ~\rsun.
We estimate the planet mass and radius to be
$\{\mpl,\rpl\}$ = \{ \koicurPPmEarth, \koicurPPrEarth \}. 
The planet's density is near \koicurPPrhoshort ~\gcmc;
it is thus slightly denser and more massive than Neptune,
but about the same size.

\end{abstract}

\section{Introduction}

Transiting extrasolar planets provide unparalleled opportunities for
detailed study of the physical characteristics of distant solar systems.
Since the first transiting planet detection a decade ago,
ground-based surveys such as TrES, HAT, and Super-WASP \citep{alo04b,
bak04, pol06} and the spaceborne telescope CoRoT
\citep{bag07} have located more than 50 transiting planets,
spanning a large range in size and mass. 
With the advent of NASA's {\it Kepler Mission}, we have a new and extraordinarily
sensitive tool for studying transiting planets.
{\it Kepler}
has enough sensitivity to detect Earth-size planets orbiting in the
habitable zones of Sun-like stars during its planned 3.5-year mission
\citep{koc10, bor10}. 
The first science data to return after {\it Kepler's} launch in Mar 2009 were time
series from a 9.7-day commissioning run, followed after a 1.6-day gap 
by a 33.5-day science run.
Here we describe the transiting planet Kepler-4b, one of several
transiting planets discovered during 
these first two observing intervals.

\section{Observations, Analysis, Tests for False Positives}

Observations of the {\it Kepler} target field commenced 1 May 2009;
the data that we describe here are long cadence (LC) photometry,
which correspond to integration times of 29.426 minutes.
For a full description of the {\it Kepler} field of view, observing modes,
and data processing pipeline, see \citet{jen10, cal10}.
For {\it Kepler} target stars brighter than $r = 13$, the RMS photometric precision
attained for relative flux time series is typically better than 
$2 \times 10^{-4}$ per 29-minute integration \citep{gill10}.
We detrended photometry from the mission data reduction 
pipeline and searched it for significant
transit-like events using the procedures described by \citet{jen10} and
by \citet{bat10}.

One of the transiting planet candidates identified by the process 
just described was the star we now term Kepler-4.
This star is uncommonly bright by {\it Kepler} standards.
Its Kepler magnitude (AB magnitude averaged over the $Kepler$ bandpass) is
${\rm Kepmag} = \koicurCCkicr$.
Characteristics of this star
are given in Table 1;  briefly, it is a somewhat metal-rich
($[Fe/H] = \koicurSMEfeh$) star of nearly solar temperature
($\koicurSMEteff$ K), evidently seen near the end of its
main-sequence lifetime, as explained below.

\begin{figure}
\plotone{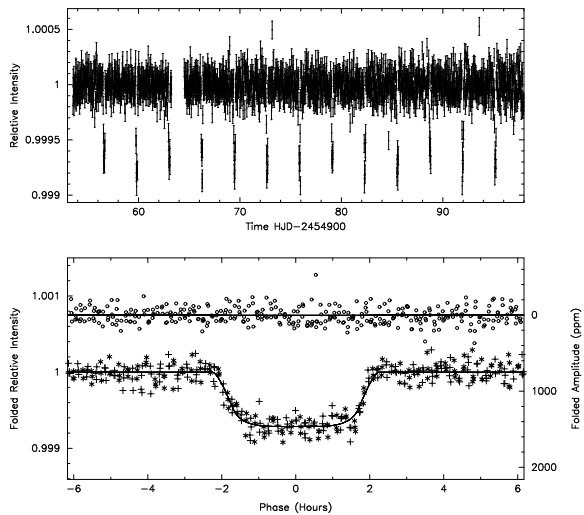}
\caption{
The phased light curve of Kepler-4b containing 13 transits observed
by the {\it Kepler} photometer between 1 May 2009 and 15 Jun 2009.
The upper panel shows the full 44-day time series after detrending.
The bottom panel shows the light curve folded with the orbital period;
different plot symbols denote odd- and even-numbered transits.
The lower curve shows transit data overplotted on the fitted transit
model.
The upper curve covers the expected time of occultation,
with the fitted (essentially constant) model overplotted.
This model assumes a circular orbit.
The full folded light curve (not shown here) gives no evidence for an
occultation at any phase.
For transit and orbital parameters, see Table 1.
\label{fig1}}
\end{figure}

Figure 1 shows the light curve for Kepler-4, folded with a period
of $\koicurLCPprec$ d.\footnote{Time series of the photometry and
of radial velocity data presented here may be retrieved from the
MAST/HLSP data archive at http://archive.stsci.edu/prepds/kepler\_hlsp.}
Since the transit signal was clear, we proceeded with follow-up
observations as described in \citet{gau10, bat10}.
Experience with both ground- and space-based transit observations shows that
a fairly large fraction of events that resemble planetary transits are
actually caused by eclipses involving only stars,
or eclipses with properties that are significantly confused by diluting
light from one or more stellar companions, either physical or projected.
For this reason, we designed our follow-up observations
to determine whether the light 
curve dips can be ascribed to a
transiting planet; if not, what eclipsing star or other process
might be responsible for them; and if so,
what the properties of the host star and its planet might be.

Ground-based visible-light speckle imaging from the WIYN Telescope,
NIR adaptive-optics imaging from the Mt. Palomar 5m telescope,
and also with the NIRC2 camera on the Keck telescope
all show Kepler-4 as an isolated star,
apart from a neighbor that is 3.5 magnitudes fainter at a distance
of 11.9$\arcsec$.
We estimate that this star contributes 2$\pm$2\% to the flux we measure
for Kepler-4, which dilutes the transit by a factor of 1.02$\pm$0.02.
We take this dilution and its uncertainty into account in computing
the the planetary radius $\rpl$ and its probable errors.
Limits on nearby companions are described by \citet{bat10};
the imagery rules out background eclipsing binaries that might
simulate the observed transits, up to a magnitude difference of 9.8
in H band, for companions between 0.12 and 3.0 arcsec from Kepler-4.
We have also searched for motion of the image centroid that is
correlated with the transits, finding no such motion
with a limit of $4 \times 10^{-4}$ arcsec.

Two reconnaissance spectra obtained with the TRES
spectrograph on the 1.5-m Tillinghast Reflector at the Whipple Observatory
showed a velocity variation of less than 150 m/s over five days.
Accordingly we obtained radial velocity measurements with the
HIRES spectrograph on the Keck I telescope \citep{vog94}.
Figure 2 shows the radial velocity data for Kepler-4, folded with
the transit period and shifted so that transit center occurs at
zero phase.
Assuming zero eccentricity,
we obtained a radial velocity variation with phase that is consistent 
with the observed light curve, with a reflex velocity amplitude $K$ of
$\koicurRVK \ \ms$ and velocity residuals of only
3.6 $\ms$.
We also performed a fit in which we allowed the eccentricity $e$ to float.
This gave $e = 0.22 \pm 0.08$ and $K = 10.0$ ms$^{-1}$; 
the reduced $\chi^2$ improved only slightly.
Finally, a Monte Carlo bootstrap (described below) that simultaneously fits all
photometry and RV values, and that is consistent with stellar evolution
models, gives $e \sin \omega = -0.003 \pm 0.058$ and
$e \cos \omega = -0.005 \pm 0.057$. 
There is thus only marginal evidence for a noncircular orbit.
In the analysis below, we do however account throughout for the possible
range of $e$ and $\omega$ in our estimates of stellar and planetary
parameters and their uncertainties.
Given the integration times of the spectra and
Kepler-4's
brightness, $\teff$, and slow rotation ($v \sin i = \koicurSMEvsin \ \kms$),
the residuals are consistent with expectations.
We also used the HIRES spectra to search for variations in the 
shapes of line bisectors;
the component of the bisector span that is in phase with the orbital
period has amplitude $-1.2 \pm 4.2 \ \ms$ (with uncertainty estimated from
the bisector scatter), offering no support for the
presence of a blended eclipsing binary star.

We can reject many photometric blend scenarios, but not all.
Kepler-4b's long transit duration rules out hierarchical triple systems
in which the primary of the eclipsing pair contributes less than
about 25\% of the system luminosity.
Near-twin star systems permitted by this constraint would lead to errors in the
planet radius and mass by factors of a few -- these would be serious, 
but not large enough 
that transits by stars
could be confused with those by planets.
Moreover, virtually all transiting planet parameter estimates 
are vulnerable to confusion
of this sort.
An unresolved background eclipsing binary star might accurately simulate
the transit light curve, but such a system would not likely produce the very
small observed RV variation, nor the small line bisector variation.
The most plausible remaining blend scenario involves an unresolved
background transiting Jupiter-size planet.
Such a blend could be consistent with all current observations, though
the required coincidence in apparent position is very unlikely a priori.
  
\begin{figure}
\plotone{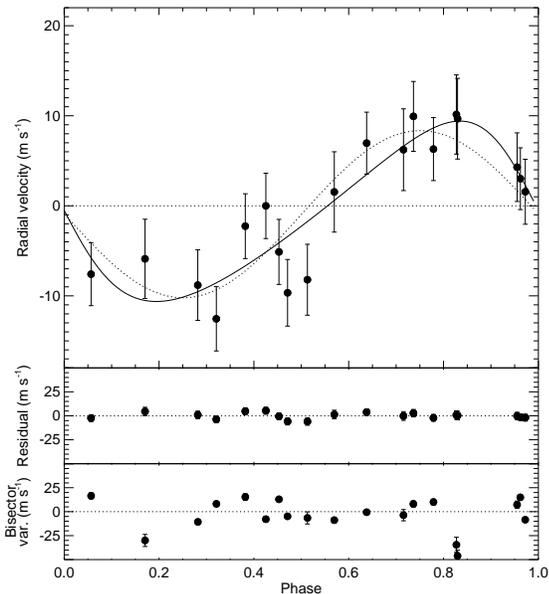}
\caption{
Top panel: The phased radial velocity curve of Kepler-4, consisting of 19 epochs observed
using the Keck/HIRES spectrometer, spanning 69 days.
The dashed overplotted curve is a fit assuming a circular orbit, phased to match the
transit photometry.
The solid curve shows the best-fit eccentric orbit, with $e=0.22$.
Middle panel: O-C residuals of velocities relative to the circular orbit fit.
Bottom panel: Bisector span for each epoch, measured between the 40\% and 90\%
depth points in the cross-correlation against a representative spectrum. 
\label{fig2}}
\end{figure}

Thus, although some kinds of confusing photometric blends are possible,
there is no evidence for them.  
In what follows, we assume
that the observed light curve
results from transits by a rather small extrasolar planet across the
face of a normal, single, Sun-like star.

\section{Properties of the Planet and Host Star}

\subsection{Method for Estimating Host Star Masses and Radii}

The {\it Kepler} photometer produces a light curve that implies the planetary radius
in units of the host star radius.
Likewise, groundbased followup Doppler spectroscopy yields the mass of the
planet in terms of the  host star's mass.
Thus, having accurate values for the host star's fundamental parameters --
mass, radius, effective temperature, and composition --  will be essential 
to carrying out {\it Kepler's}
mission to characterize planets circling distant stars.

But it is also
possible to reverse this logical flow, and use transiting planets
as probes of their host stars.
\citet{sea03} 
derived expressions relating the light curve to the mean density $\rho_*$ 
of a transiting planet's
host star.
They showed that $\rho_*$ can be expressed in terms of the planet's orbital
period, the fractional flux obstructed by the planet, and
two different measures of the transit duration.
All of these quantities are directly measurable from the light curve,
and their interpretation depends only on Newtonian mechanics and geometry.

The method we use (henceforth, the $\rho_*$ method) 
for estimating stellar radii and other parameters
is an implementation of that described by \citet{soz07} 
and subsequently 
used by, e.g., \citet{bak07, win07} and \citet{cha07}.
This technique has become the most trusted way of obtaining information
about transiting planets' host stars, as illustrated by \citet{tor08}.

The basis for the $\rho_*$ method is that, 
although stellar models are ordinarily taken to depend on 5 parameters
\{mass, age, metallicity, initial helium abundance, and mixing length\} = 
$\{ M_*, A_*, [Z], Y_0, \alpha \}$,
the last two of these do not vary much among stars, and
for practical purposes are often taken as known.
In particular, the much-used Yonsei-Yale (henceforth YY) model grid 
\citep{yi01} uses specified values
for the mixing length and for the initial helium abundance.
In this approximation, stars are described by a 3-dimensional grid of models,
parameterized by mass, age, and metallicity.
Every transiting planet
discovered by {\it Kepler} and followed up with reconnaisance spectroscopy
has available a transit light curve that constrains $\rho_*$,
and also spectrographic estimates of $\teff$ and $[Z]$.
Moreover, for main-sequence stars, the problem of inferring stellar structure
parameters from the 3 observables
$\{\rho_*,\ \teff,\ [Z]\}$ is usually well-posed, allowing precise
conclusions without important degeneracies.
Thus, the quantities that we may readily observe usually suffice to isolate
a single set of model parameters \{mass, age, metallicity\};
from these, we may compute any other global property (e.g. radius, luminosity,
or surface gravity).

We implement the $\rho_*$ method by searching a 
precomputed grid of models
(interpolated from the YY models)
to find the one that best matches the observations in a $\chi^2$ sense.
The optimization problem involves interpolating within the given model
grid and performing a conventional (eg ``amoeba'') hill-climbing optimization
search.
Estimating errors and verifying that the search has converged to the
global optimum is done using a
Markov Chain Monte Carlo (MCMC) process.
To allow for possible systematic errors in the spectrographic analysis,
we took uncertainties in $[Z]$ to be the larger of 
0.06 dex and the quoted value,
and  twice the quoted value for $\teff$.
For $\rho_*$, we used a parameterization of the (often highly asymmetric) error
distribution estimated from a jack-knife analysis of the photometry.
Not all parameter choices $\{ \mstar, A_*, [Z] \}$ permitted by a given
$\rho_*$ are consistent with the photometry, and vice versa.
Moreover, in the light curve analysis, the estimated $\rho_*$ depends somewhat
on the initial guess for $\mstar$ \citep{koc10}
and on the range of orbital eccentricity allowed by the observations.
To account for the different information supplied by these two kinds
of model fitting, we iterated the solution by putting the MCMC probability
distributions back into the light curve analysis.
The most likely parameter values and uncertainty distributions we report
below are the result of this once-iterated fit.

\citet{bro09} has tested the $\rho_*$ method against a sample of 169 stars
(mostly members of eclipsing binaries taken from the compilation by
\citet{tor09}) that have accurately known masses
and radii.
This comparison
shows that the method's systematic errors do not exceed 2-3\%
in radius, and about 6-9\% in mass.
Random errors appear to arise about equally from uncertainties
in the estimated stellar $\teff$ and metallicity, and from uncertainties
in the masses and radii inferred for the eclipsing binary components
using traditional (but largely model-independent) techniques.
Systematic errors arise mostly because rapidly-rotating and hence 
magnetically active stars are seen to have larger radii than slowly-rotating
stars with otherwise similar properties (eg, \citet{tor06, lop07, mor08}).
The sample of eclipsing binary stars consists almost entirely of such
fast rotators;
this causes an excess of up to several percent in the actual radii of stars
of sub-solar mass, relative to the YY models.
On the other hand,
the host stars of confirmed transiting planets found by {\it Kepler} 
are mostly slow rotators (if only because it is difficult to measure precise
radial velocities of rapidly-rotating stars).
Errors in these estimates should therefore be dominated by
measurement errors, especially in $[Z]$ and in $\rho_*$.

\subsection{The Star Kepler-4}

Simultaneous fits to the transits of Kepler-4b and to the RV data 
(allowing nonzero eccentricity) yield 
the ratio of orbital
semimajor axis $a$ to stellar radius $R_*$, namely $a/\rstar = 
\koicurLCar$, 
which with the orbital period
gives $\rho_* = 0.50\pm0.16 \ \gcmc$.
This is a low value, roughly 1/3 that of the Sun, 
suggesting that Kepler-4 must have evolved
considerably away from the zero-age main sequence.
Applying the $\rho_*$ method confirms this conclusion, and in fact
finds a range of model parameters, 
all of which fit the observations
almost equally well, but which imply distinct evolutionary
states for the star.
Figure 3 shows evolution tracks for three models having
masses between 1.13 $\msun$ and 1.28 $\msun$.
In this mass range, old main-sequence stars have small convective cores.
Because of efficient mixing, these cores suffer hydrogen exhaustion
all at once, following which the entire star must compress and heat
until the central temperature rises enough to start shell hydrogen
burning.
The blue-going hook at main-sequence turnoff is a result of this
compression.
In the case of Kepler-4, the observed $\rho_*$ and $\teff$
(indicated by the error box in Figure 3) are closely matched by
each of the evolution tracks, albeit at different ages and with
models occupying different evolutionary states.
Thus, for the lowest-mass star to fit the observations, it must have
just finished its core-contraction phase, and be on its way to becoming
a shell-burning subgiant.
The highest-mass star to fit the same observations must
still be on the main sequence but nearing hydrogen exhaustion 
in its core,
just poised to begin core contraction.

\begin{figure}
\plotone{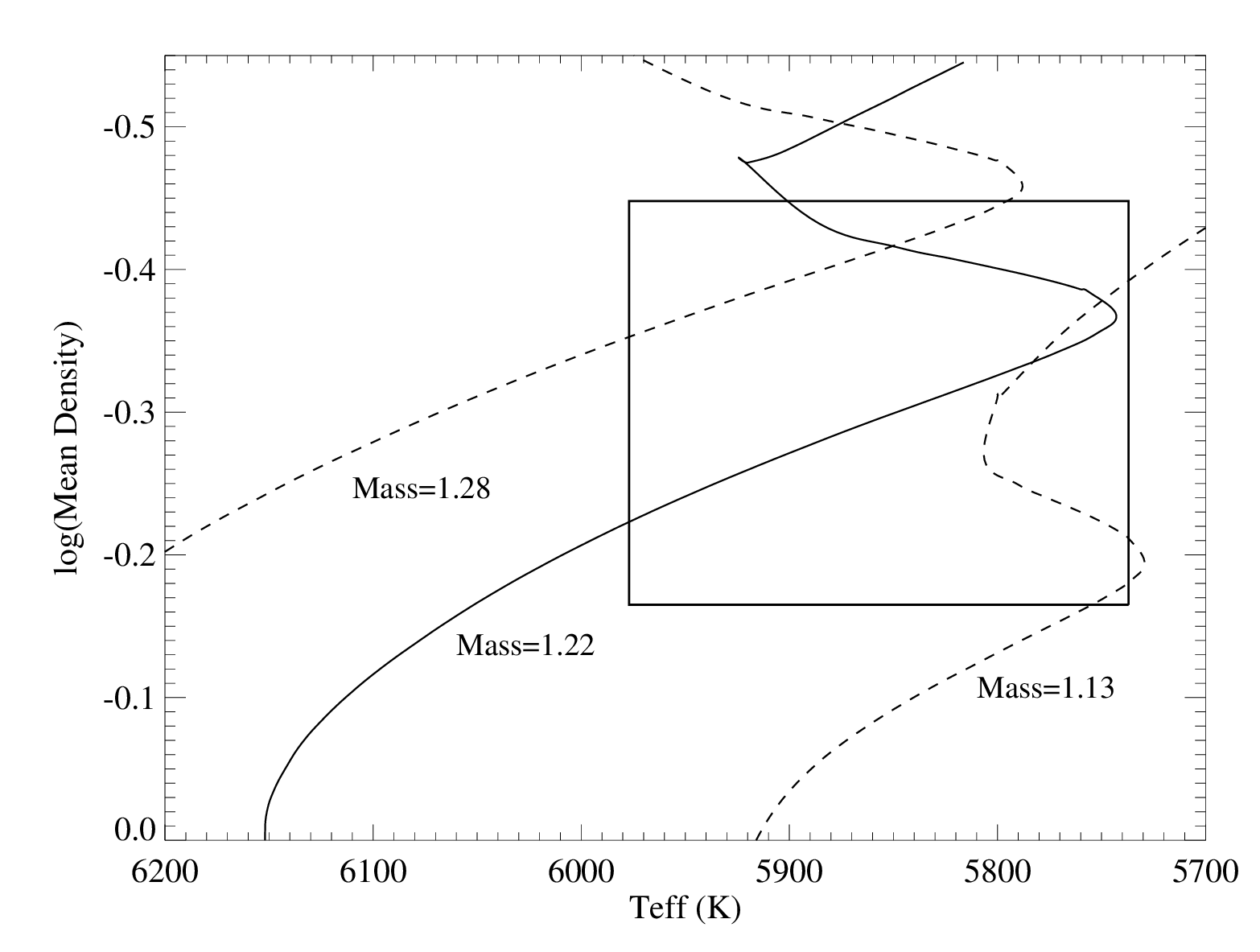}
\caption{
Yonsei-Yale evolutionary tracks for stars with $[Z] = 0.17$
and masses between 1.13 and 1.28
$\msun$, plotted in the $\rho_* - \teff$ plane.
The box indicates our adopted observational constraints on $\rho_*$ and $\teff$.
For this range of masses, all tracks lie partly inside the error box, showing
that multiple solutions are possible that satisfy these constraints.
\label{fig3}}
\end{figure}

It is not possible to distinguish among these possibilities using existing
observations.
Although the masses of acceptable models differ by as much as 13\%, the radii
differ by only 1/3 as much (in order to give the same observed mean density).
This radius difference would cause a difference in $\log(g)$ that is too
small to measure with current methods.
Likewise, the luminosity difference of about 8\% implies a parallax
difference of only 4\%, also too small to discern at this star's likely distance
of about \koicurXdist ~pc.
Asteroseismology may however offer a way out of this quandry.
Because of the star's relatively large luminosity and apparent brightness,
it may be possible to measure the frequencies of its pulsation modes,
and hence distinguish among the feasible evolution scenarios.
Kepler-4 is now being observed with {\it Kepler's} short (60 s) cadence;
the mission will report pulsation properties estimated from these data
if and when the pulsation signals rise above the noise.

We list the inferred properties of Kepler-4
in Table 1.
For the quantities $\mstar$ and $\rstar$ 
we give uncertainties that include the effects of observational
uncertainties, of uncertainty in orbital eccentricity, and of
uncertainty in the host star's evolutionary state.

\begin{figure}
\plotone{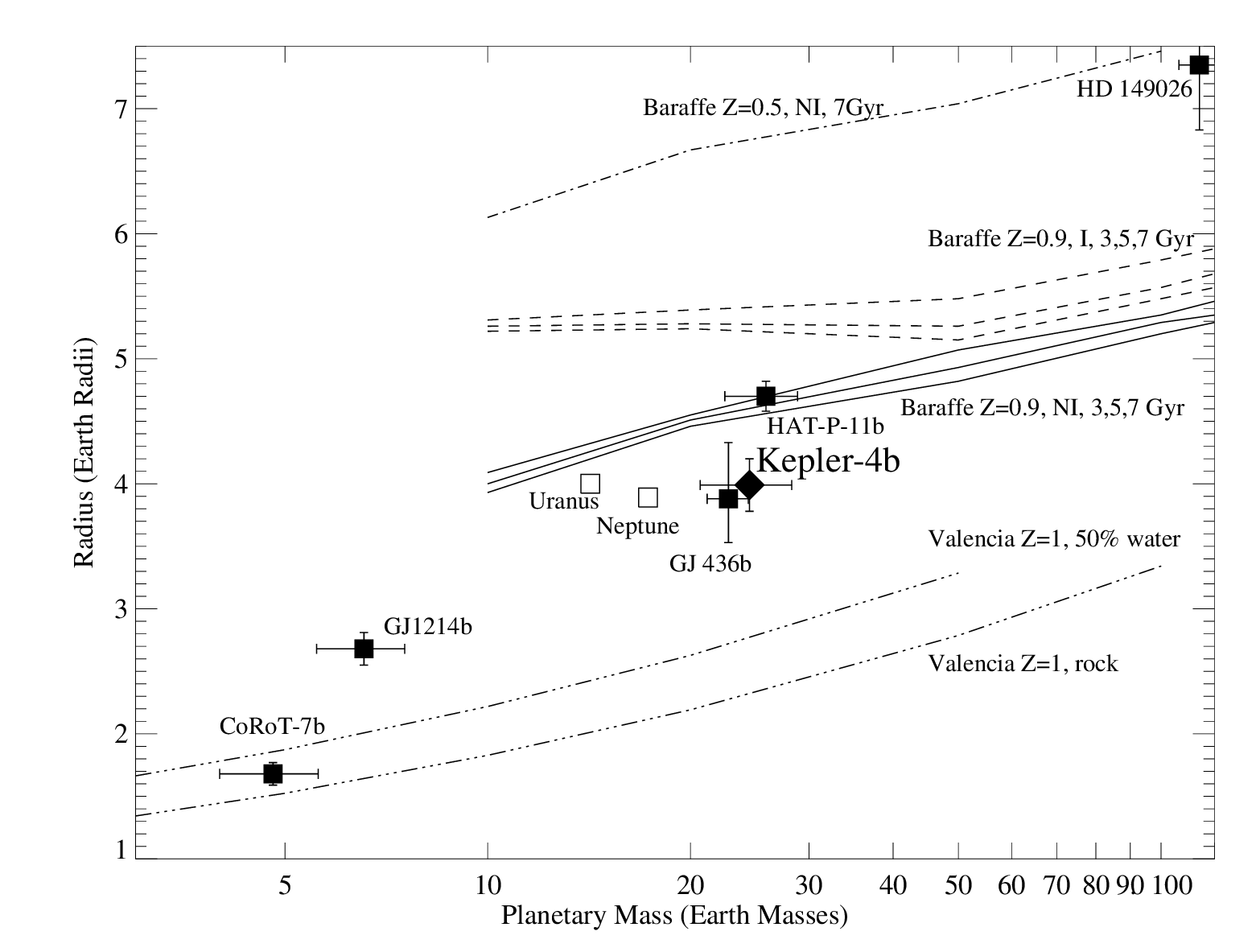}
\caption{
Kepler-4b is shown in a mass-radius plot, along with other hot Neptunes,
the hot super-Earths CoRoT-7b \citep{leg09, que09} and GJ1214b \citep{cha09}, 
and
model curves by \citet{bar08} and by \citet{val07}, showing increasing
heavy-element fractions toward the bottom of the Figure.
Multiple dashed and solid curves show model results for differing ages;
``I'' indicates irradiated and ``NI'' non-irradiated models.
Kepler-4b appears to be denser than HAT-P-11b but similar to GJ 436b,
denser than the non-irradiated $Z = 0.9$ model by \citet{bar08},
but much less dense than is expected for a water or rocky planet.
\label{fig4}}
\end{figure}

\subsection{The Planet Kepler-4b}

Given the properties of its host star, the depth of transits due to Kepler-4b
implies a planetary radius of \koicurPPrlong $\rjup$ = \koicurPPrEarth,
and a mass of \koicurPPmlong  $\mjup$ = \koicurPPmEarth.
As with $\mstar$ and $\rstar$, these values and uncertainties should be
interpreted as describing the centers and 68\%-probability points 
of the marginal probability distributions, accounting for all of the
uncertainties.
Kepler-4b is therefore slightly more massive than
Neptune, and about the same size.
Its mean density is about \koicurPPrhoshort $\ \gcmc$, greater
than Jupiter's and considerably larger than Saturn's.
Assuming a 10\% Bond albedo and efficient heat redistribution to the planet's
night side, its equilibrium temperature is \koicurPPteq  K \citep{koc10}.
Radial velocity data provide
no evidence for other massive
planets in small orbits.
Because of the limited time span and precision of the extant RV observations,
it is however impossible to rule out such planetary companions.

\section{Discussion}

Figure 4 shows Kepler-4 in a mass-radius diagram that spans the range so far
occupied by small transiting exoplanets.
Kepler-4b is the third known transiting Neptune-like planet, 
together with GJ436b and HAT-P-11b; 
all three have masses and radii equal or larger than Neptune and Uranus. 
The most important differences between the 3 exoplanets are their host stars and 
the incident flux the planets receive 
(GJ436b: M dwarf, $\teq$ = 650 K; HAT-P-11b: K dwarf, 880 K; Kepler-4b: G subgiant, 1650 K). 
On the mass-radius diagram Kepler-4b and GJ436b both lie below the 
$Z=0.9$ Baraffe et al.(2008) non-irradiated model. In these models, $Z=0.9$ denotes that the planet
has a 10\% by mass envelope of H and He, and 'non-irradiated' corresponds roughly the case of 
GJ436b and cooler. The fact that Kepler-4b and GJ436b have essentially identical radii
(3.88 $\pm$ 0.15 $R_{\earth}$ for GJ436b from NASA EPOXI 
\citep{bal09} as compared to 3.99 $\pm$ 0.21 $R_{\earth}$ for Kepler-4) at the same mass,
despite Kepler-4b's high $\teq$, implies a difference in bulk composition. We suggest that
Kepler-4b has a H/He envelope of about 4-6\% by mass and a correspondingly higher water and rock 
fraction. However, the inherent degeneracy in this part of the mass-radius diagram means that
the H/He envelope might be slightly more massive if the rock-to-water ratio in the interior of
Kepler 4b is unusually high (i.e., significantly higher than in Neptune or Uranus).
Nevertheless, we can state with a measure of confidence that there are no possible interior
models for Kepler-4b with no H/He envelope and neither it nor GJ436b is 
compact enough to be a water-rich super-Earth.

The slightly eccentric orbits of the two previously known transiting Neptune-like planets, 
in similar potentials and at similar ages, suggest that their interiors are 
somewhat less dissipative to stellar tides than the giants in our Solar System. 
Therefore, it would be very interesting to resolve whether the orbit of Kepler-4b 
has comparable eccentricity or is circular; 
the current observations are still inconclusive.

\acknowledgments

Funding for this Discovery mission is provided by NASA's Science Mission Directorate.
We are grateful first to the entire {\it Kepler} team, past and present.
Their tireless efforts were all essential to the success of the mission.
For special advice and assistance, we thank Lars Buchhave, David Ciardi, 
Megan Crane, Willie Torres,
Mike Haas, and Riley Duran.
{\it Facilities:} \facility{The Kepler Mission}.

\begin{deluxetable}{lcc}
\tabletypesize{\scriptsize}
\tablewidth{0pc}
\tablenum{1}
\tablecaption{System Parameters for \koicur \label{tab:parameters}}
\tablehead{\colhead{Parameter}  & 
\colhead{Value}                 &
\colhead{Notes}}
\startdata
\sidehead{\em Transit and orbital parameters}
Orbital period $P$ (d)                          & \koicurLCP            & A     \\
Midtransit time $E$ (HJD)                       & \koicurLCT            & A     \\
Scaled semimajor axis $a/\rstar$                & \koicurLCar           & A     \\
Scaled planet radius \rpl/\rstar                & \koicurLCrprstar      & A     \\
Impact parameter $b \equiv a \cos{i}/\rstar$    & \koicurLCimp          & A     \\
Orbital inclination $i$ (deg)                   & \koicurLCi            & A     \\
Orbital semi-amplitude $K$ (\ms)                & \koicurRVK            & A,B     \\
Orbital eccentricity $e$                        & 0 (adopted)           & A,B     \\
Center-of-mass velocity $\gamma$ (\ms)          & \koicurRVgamma        & A,B     \\
\sidehead{\em Observed stellar parameters}
Effective temperature \teff\ (K)                & \koicurSMEteff        & C     \\
Spectroscopic gravity \logg\ (cgs)              & \koicurSMElogg        & C     \\
Metallicity \feh                                & \koicurSMEfeh         & C     \\
Projected rotation \vsini\ (\kms)               & \koicurSMEvsin        & C     \\
Mean radial velocity (\kms)                     & \koicurRVmean         & B     \\
\sidehead{\em Derived stellar parameters}
Mass \mstar (\msun)                             & \koicurYYmlong        & C,D     \\
Radius \rstar (\rsun)                           & \koicurYYrlong        & C,D     \\
Surface gravity \loggstar\ (cgs)                & \koicurYYlogg         & C,D     \\
Luminosity \lstar\ (\lsun)                      & \koicurYYlum          & C,D     \\
Absolute V magnitude $M_V$ (mag)                & \koicurYYmv           & D     \\
Age (Gyr)                                       & \koicurYYage          & C,D     \\
Distance (pc)                                   & \koicurXdist          & D     \\
\sidehead{\em Planetary parameters}
Mass \mpl\ (\mjup)                              & \koicurPPm            & A.B,C,D     \\ 
Radius \rpl\ (\rjup, equatorial)                & \koicurPPr            & A.B,C,D   \\ 
Density \rhopl\ (\gcmc)                         & \koicurPPrho          & A,B,C,D     \\
Surface gravity \loggpl\ (cgs)                  & \koicurPPlogg         & A,B,C,D     \\
Orbital semimajor axis$a$ (AU)                  & \koicurPParel         & E     \\
Equilibrium temperature \teq\ (K)               & \koicurPPteq          & F 
\enddata
\tablecomments{\\
A: Based primarily on the photometry.\\
B: Based on the radial velocities.\\
C: Based on spectrum analysis (FIES/MOOG or HIRES/SME).\\
D: Based on the Yale-Yonsei evolution tracks.\\
E: Based on Newton's version of Kepler's Third Law.\\
F: Assumes Bond albedo = 0.1 and complete redistribution.\\
}
\end{deluxetable}

\end{document}